\documentclass[12pt]{article}
\usepackage{fullpage}
\usepackage{graphicx}
\newcommand{\be}{\begin{equation}}
\newcommand{\ee}{\end{equation}}
\newcommand{\bphi}{\mbox{\boldmath $\phi$}}

\def\w {\omega}
\font\mybb=msbm10 at 12pt

\def\bb#1{\hbox{\mybb#1}}

\def\bZ {\bb{Z}}

\newcommand{\news}{\setcounter{equation}{0}\quad}
\def\ben{\begin{equation}}
\def\een{\end{equation}}
\def\bea{\begin{eqnarray}}
\def\eea{\end{eqnarray}}
\begin{document}
\title{
\begin{flushright}\ \vskip -2cm {\small{\em DCPT-09/07}}\end{flushright}\vskip 2cm 
Baby Skyrmions stabilized by vector mesons}
\author{David Foster and Paul Sutcliffe\\[10pt]
{\em \normalsize Department of Mathematical Sciences,
Durham University, Durham DH1 3LE, U.K.}\\[10pt]
{\normalsize Emails:  d.j.foster@durham.ac.uk, 
\quad p.m.sutcliffe@durham.ac.uk}
}

\date{January 2009}
\maketitle
\begin{abstract}
Recent results suggest that multi-Skyrmions stabilized by $\omega$ mesons
have very similar properties to those stabilized by the Skyrme term.
In this paper we present the results of a detailed numerical investigation
of a (2+1)-dimensional analogue of this situation. Namely, we compute
solitons in an $O(3)$ $\sigma$-model coupled to a massive
vector meson and compare the results to baby Skyrmions, which are
solitons in an $O(3)$ $\sigma$-model including a Skyrme term.
We find that multi-solitons in the vector meson model are surprisingly
similar to those in the baby Skyrme model, and we explain this
correspondence using a simple derivative expansion.
\end{abstract}

\newpage
\section{Introduction}\news
The Skyrme model \cite{Sk} is a nonlinear theory of pions in which
baryons are described by soliton solutions. With only pion degrees
of freedom the Skyrme model requires the inclusion of the Skyrme term, 
which is quartic in derivatives, in order to yield stable soliton solutions.
Many years ago it was realized that including $\omega$ mesons,
in addition to pions, produces a stable soliton solution without the
need for a Skyrme term \cite{AN2}. Unfortunately, 
numerical computations of multi-solitons
in the $\omega$ meson model is a formidable challenge, and to date
no multi-soliton solutions have been computed. This is in contrast to 
the Skyrme model, where numerical solutions have been obtained up to reasonably
large soliton numbers \cite{BS3}. However, by using analytical approximations,
recent progress has been made on the study of multi-solitons
in the $\omega$ meson model \cite{Su}, and the results suggest a 
surprising similarity with solitons in the Skyrme model.

The baby Skyrme model \cite{PSZ}
is a (2+1)-dimensional analogue of the Skyrme model.
It is a modified $O(3)$ $\sigma$-model that includes a Skyrme term.
In this paper we study the soliton solutions of an alternative to the
baby Skyrme model, in which the Skyrme term is removed and stabilization
is achieved by including a massive vector meson.
A comparison is made between the multi-solitons in the two models,
as this is a more tractable lower-dimensional analogue of the 
(3+1)-dimensional Skyrmion situation discussed above.
We find a remarkable similarity between the multi-solitons in the two
models, at both the qualitative and quantitative level,
and we explain this behaviour by using a simple derivative expansion.
These results provide a further justification for the approximate
techniques used in (3+1)-dimensions \cite{Su}, that motivated the
present work.

\section{The baby Skyrme model and vector mesons}\news
The Lagrangian density of the baby Skyrme model in (2+1)-dimensions
is given by \cite{PSZ}
\be
{\cal L}^{\rm BS}=\frac{1}{2}\partial_\mu\bphi\cdot\partial^\mu\bphi
-m^2(1-\phi_3)
-\frac{\kappa^2}{4}(\partial_\mu\bphi\times\partial_\nu\bphi)\cdot
(\partial^\mu\bphi\times\partial^\nu\bphi),
\label{bslag}
\ee
where $\bphi=(\phi_1,\phi_2,\phi_3)$ is a three-component unit
vector $\bphi\cdot\bphi=1,$ and $m$ is the mass of the $\phi_1$ and
$\phi_2$ fields, which are the analogues of the pions fields in
the Skyrme model.
The final term in (\ref{bslag}) is the Skyrme term, with $\kappa$ a 
positive constant. The size of the baby Skyrmion is determined by
the combination $\sqrt{\kappa/m}.$ 
 
Finite energy requires the boundary condition
$\bphi\rightarrow{\bf e}_3=(0,0,1)$ as $|{\bf x}|\rightarrow\infty,$ which
implies that the spatial plane is compactified to the two-sphere.
Therefore, at fixed time, $\bphi: S^2\mapsto S^2,$ with an associated
integer winding number, $B\in\bZ=\pi_2(S^2),$ which is the soliton number. 

There is a conserved topological current
\be
B^\mu    = -\frac{1}{8\pi}\varepsilon^{\mu\alpha\beta}\bphi
\cdot (\partial_\alpha\bphi\times\partial_\beta\bphi).
\label{topcur}
\ee
and the soliton number is the integral over space of the topological
charge density $B^0,$ that is, $B=\int B^0\, d^2x.$

Baby Skyrmion solutions for $1\le B \le 6$ were computed numerically in 
\cite{PSZ}, using the parameter values $m=1/\sqrt{10}$ and $\kappa=1.$ 
Briefly, the solitons are all bound
states and are axially symmetric
for $B\le 2,$ but have only discrete symmetries for $B>2.$
We shall discuss these solutions in more detail in the following
Section.

Our alternative to the baby Skyrme model is given by the Lagrangian
density
\be
{\cal L}^{\rm VM}=\frac{1}{2}\partial_\mu\bphi\cdot\partial^\mu\bphi
-m^2(1-\phi_3)-\frac{1}{4}(\partial_\mu\w_\nu-\partial_\nu\w_\mu)
(\partial^\mu\w^\nu-\partial^\nu\w^\mu)
+\frac{1}{2}M^2\w_\mu\w^\mu+g\w_\mu B^\mu,
\label{vmlag}
\ee
where the Skyrme term has been removed and a vector field $\omega_\mu$
with mass $M$ has been added. This is the analogue of the $\omega$ 
meson in the (3+1)-dimensional theory, hence the notation. 
The coupling of the vector field to the topological current, with
positive coupling constant $g,$ also mirrors the higher dimensional
theory \cite{AN2}.

In the remainder of this paper we are concerned with static solutions of the
vector meson theory (\ref{vmlag}). For static fields the spatial
components of the topological current vanish $B^i=0$ and therefore
$\w_i=0,$ since the topological current $B^\mu$ provides the source for
$\w_\mu.$

Only static fields are considered from now on, so for notational 
convenience we write $\w\equiv \w_0.$ 
With $\w_i=0,$ the static energy derived from (\ref{vmlag}) 
is given by
\be
E^{\rm VM}=\int \Bigl(\frac{1}{2}\partial_i\bphi\cdot\partial_i\bphi
+m^2(1-\phi_3)
-\frac{1}{2}\partial_i\w\partial_i\w-\frac{1}{2}M^2\w^2
+\frac{g}{8\pi}\w\epsilon_{ij}\bphi\cdot 
(\partial_i\bphi\times\partial_j\bphi)
\Bigr)\, d^2x.
\label{en}
\ee
The static field equations that follow from the variation of (\ref{en})
are
\be
\partial_i\partial_i\bphi+m^2{\bf e}_3
+\frac{g}{4\pi}\epsilon_{ij}\partial_j\omega\,\bphi\times\partial_i\bphi
+(\partial_i\bphi\cdot\partial_i\bphi-m^2\phi_3)\bphi=0,
\label{eom1}
\ee
and
\be
\partial_i\partial_i\omega-M^2\omega
=-\frac{g}{8\pi}\epsilon_{ij}\bphi\cdot (\partial_i\bphi\times\partial_j\bphi).
\label{eom2}
\ee
For fields that satisfy equation (\ref{eom2}) then multiplication
of this equation by $\omega$ and its substituion into 
(\ref{en}), together with an integration by parts, allows the energy (\ref{en})
to be rewritten in the form
\be
E^{\rm VM}=\int \Bigl(\frac{1}{2}\partial_i\bphi\cdot\partial_i\bphi
+m^2(1-\phi_3)
+\frac{g}{16\pi}\w\epsilon_{ij}\bphi\cdot 
(\partial_i\bphi\times\partial_j\bphi)
\Bigr)\, d^2x,
\label{en2}
\ee
which will be convenient later.

One way to see how the vector meson theory evades Derrick's theorem 
\cite{De}, on the non-existence of solitons, is to formally solve
(\ref{eom2}) for $\omega$ in terms of its Green's function and substitute
this back into the energy expression (\ref{en2}). 
This formulation presents the model in terms of a non-local interaction,
and is similar to planar soliton models used in the study of  
quantum Hall ferromagnets \cite{fqhe}. However, a more informative approach
is to approximate the solution of (\ref{eom2}) by applying a derivative 
expansion. In fact, only the leading order term is required, which simply
corresponds to neglecting the Laplacian term in (\ref{eom2}) so that
the approximate solution is simply
\be
\omega\approx 
\frac{g}{8\pi M^2}\epsilon_{ij}\bphi\cdot 
(\partial_i\bphi\times\partial_j\bphi).
\label{approxsoln}
\ee 
Substituting this approximation into the energy (\ref{en2}) gives
\be
E^{\rm VM}\approx\int \Bigl(\frac{1}{2}\partial_i\bphi\cdot\partial_i\bphi
+m^2(1-\phi_3)
+\frac{g^2}{32\pi^2M^2}|\partial_1\bphi\times\partial_2\bphi|^2
\Bigr)\, d^2x,
\label{approxen}
\ee
which is precisely the static energy of the baby Skyrme model
(\ref{bslag}) upon identification of the parameter $\kappa$ as
\be
\kappa=\frac{g}{4\pi M}.
\label{kappa}
\ee
This analysis suggests that soliton solutions should be 
similar in the vector meson and baby Skyrme models. In the following
Section we shall confirm this expectation by presenting the results
of numerical computations of solitons.

Note that the approximation used in the above analysis becomes more
accurate as $M$ increases, as the mass term in (\ref{eom2}) 
is increasingly dominant over the neglected Laplacian term.
In particular, the analysis is not valid in the massless case
$M=0.$ In this case equation (\ref{eom2}) reveals that $\omega$ does
not scale with a rescaling of the spatial coordinates, therefore the
interaction energy in (\ref{en2}) has the same scale invariance
as the $\sigma$-model energy. A rigorous mathematical analysis of the
existence and uniqueness properties of theories of this type 
defined on a torus can be found in \cite{St}. 
In the massless limit $M=0,$ lump-like solutions with an arbitrary scale
can only exist if $m=0,$ so that the total energy is scale invariant.
It might be interesting to investigate the lump solutions of 
such a doubly massless model, $M=m=0,$ but we shall not pursue this here.

\section{Soliton solutions}\news
As in the baby Skyrme model, we expect that for $B=1$ and $B=2$ 
the minimal energy solitons are axially symmetric.
The axially symmetric ansatz has the form $\omega(r),$
with boundary conditions $\omega'(0)=0,\ \omega(\infty)=0,$ and 
\be
\bphi=(\sin f\cos B\theta,\sin f\sin B\theta, \cos f),
\ee
where the profile function $f(r)$ has boundary conditions 
$f(0)=\pi$ and $f(\infty)=0.$

With this ansatz the static energy (\ref{en}) becomes
\be
E^{\rm VM}=2\pi\int_0^\infty\Bigl(
\frac{1}{2}f'^2+\frac{B^2}{2}\frac{\sin^2f}{r^2}+m^2(1-\cos f)
-\frac{1}{2}\w'^2-\frac{1}{2}M^2\w^2+\frac{gB}{4\pi}\w f'\frac{\sin f}{r}
\Bigr)
r\, dr,
\ee
and the static field equations (\ref{eom1}) and (\ref{eom2})
reduce to the two ordinary differential equations
\be
f''+\frac{1}{r}f'-\frac{B^2}{2r^2}\sin(2f)-m^2\sin f
+\frac{gB}{4\pi}\w'\frac{\sin f}{r}=0,
\label{ode1}
\ee
\be
\w''+\frac{1}{r}\w'-M^2\w+\frac{gB}{4\pi}f'\frac{\sin f}{r}=0.
\label{ode2}
\ee
In order to compare the vector meson model with the baby Skyrme
model we need to choose some parameter values. For the baby Skyrme
model we fix the parameters to those used in the original
investigations \cite{PSZ} and given earlier as 
$m=1/\sqrt{10}$ and $\kappa=1.$ 
Using the same numerical codes that we apply below to the vector
meson model, we have recalculated the energies
of solitons in the baby Skyrme model. For $B=1$ and $B=2$ we compute the 
baby Skyrmion energies to be $E^{\rm BS}_1=19.66$\, and 
$E^{\rm BS}_2=36.90,$ which agree with the values presented in \cite{PSZ}
to a good accuracy. 

Next we turn to choosing the parameters in the vector meson
model. Motivated by the higher-dimensional theory, we choose $M$ so that
the ratio $m/M$ is of a similar order to the ratio of the pion to
$\omega$ meson mass. The value $M=3/2$ is reasonable from this point of
view and we take this from now on. Given the earlier comments, the expectation
is that the results will not be too sensitive to the value of $M,$ providing
it is sufficiently large. 

The only remaining parameter is $g.$ 
Given the values of the other parameters, the formula
(\ref{kappa}) suggests $g=4\pi M \kappa =6\pi\approx 18.85.$
However, rather than using this value, which relied upon the
approximation (\ref{approxsoln}), we fix $g$ by requiring that the 
energy of the $B=1$ soliton is the same as in the baby Skyrme model.
This results in the slightly larger value $g=20.83,$ which we shall
use from now on.

Solving the ordinary differential equations (\ref{ode1}) and
(\ref{ode2}) using a heat flow method yields the energies
$E^{\rm VM}_1=19.66$ and $E^{\rm VM}_2=37.32<2E^{\rm VM}_1.$ 
By construction $E^{\rm VM}_1=E^{\rm BS}_1,$ but it is a non-trivial
result that $E^{\rm VM}_2$ and $E^{\rm BS}_2$ are very close:
Table~\ref{tab-energy} lists the energies in the two models 
for ease of comparison.
\begin{table}[ht]
\centering
\begin{tabular}{|c|c|c|}
\hline
$B$ & $E^{{\rm BS}}$ & $E^{{\rm VM}}$ \\
\hline
1 & 19.66 & 19.66\\
2 & 36.90 & 37.32\\
3 & 55.58 & 56.19\\
4 & 73.61 & 74.48\\
\hline
\end{tabular}
\caption{Energies of the minimal energy solitons with $1\le B\le 4.\ $
Here $E^{{\rm BS}}$ is the energy in the baby Skyrme model 
and $E^{{\rm VM}}$ is the energy in the vector meson model.
}
 \label{tab-energy}
\end{table}

\begin{figure}
\begin{center}
\includegraphics[width=11cm]{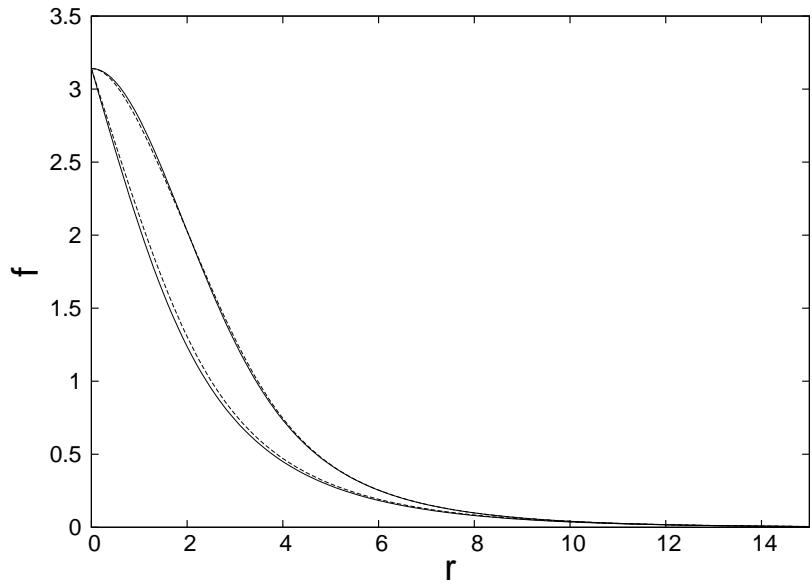}
\caption{Comparison of the profile functions for $B=1$ 
and $B=2$ solitons in the
vector meson model (solid curves) and the baby Skyrme model (dashed curves).
The lower curves correspond to $B=1$ and the upper curves to $B=2.$
}
\label{fig-cfskom}
\end{center}
\end{figure}

Figure~\ref{fig-cfskom} presents the associated profile functions
for $B=1$ and $B=2,$
with the two solid curves being those of the vector meson model
and the two dashed curves those of the baby Skyrme model.
This demonstrates the remarkable similarity between the soliton
solutions of the two models, and goes beyond the agreement of the
energies. 

As in the baby Skyrme model \cite{PSZ}, it turns out that 
in the vector meson model the minimal 
energy solitons for $B>2$ are not axially symmetric. In fact, by computing
the energies of axially symmetric solutions, it is already clear that
there is no axially symmetric bound state with $B=4$ since it has
an energy of $77.63>2E^{\rm VM}_2.$  

\begin{figure}
\begin{center}
\includegraphics[width=11cm]{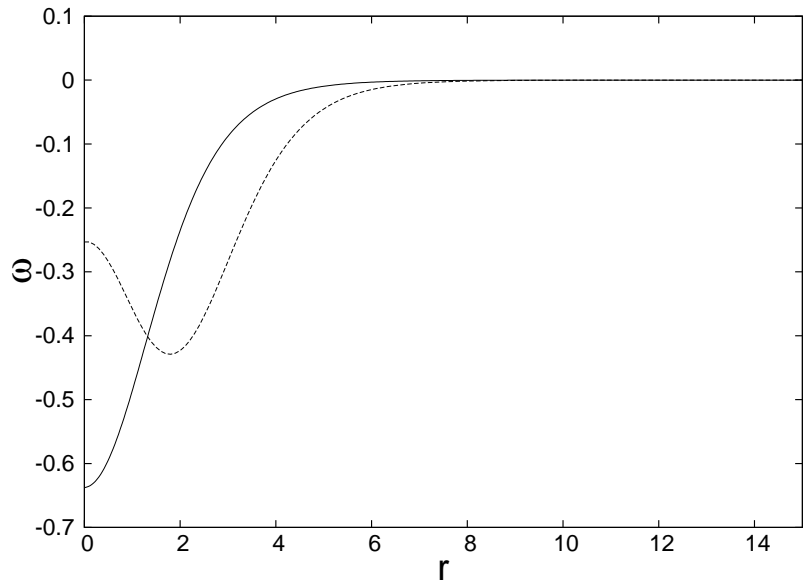}
\caption{
The axially symmetric field $\omega(r)$ for solitons
in the vector meson model with
$B=1$ (solid curve) and $B=2$ (dashed curve).
}
\label{fig-om12}
\end{center}
\end{figure}


In order to study solitons with $B>2,$ numerical solutions of the
full two-dimensional static field equations (\ref{eom1}) and 
(\ref{eom2}) must be computed. The numerical algorithm uses a heat flow
method applied to these equations, with spatial derivatives approximated
by fourth-order accurate finite differences with a lattice spacing
$\Delta x=0.2$ and a grid containing $200\times 200$ lattice points.

For comparison, we also compute the soliton solutions of the baby
Skyrme model using the same code. For $B=1$ and $B=2$ the energies
calculated using the two-dimensional code agree with those of the axially 
symmetric computations to the accuracy presented in Table~\ref{tab-energy}.
Note that our energy values for $B>2$ differ slightly from those presented
in \cite{PSZ}. Comparing with the axially symmetric computations for $B=1$ and 
$B=2,$ reveals that the two-dimensional results of \cite{PSZ} 
under-estimate the energy by approximately $1\%,$ whereas our values agree to 
a greater accuracy. This improvement is probably 
a result of our use of fourth-order accurate finite differences, since we
also find a similar under-estimate if we use only second-order accurate
finite differences.  
  
\begin{figure}[ht]
\begin{center}
\includegraphics[width=11cm]{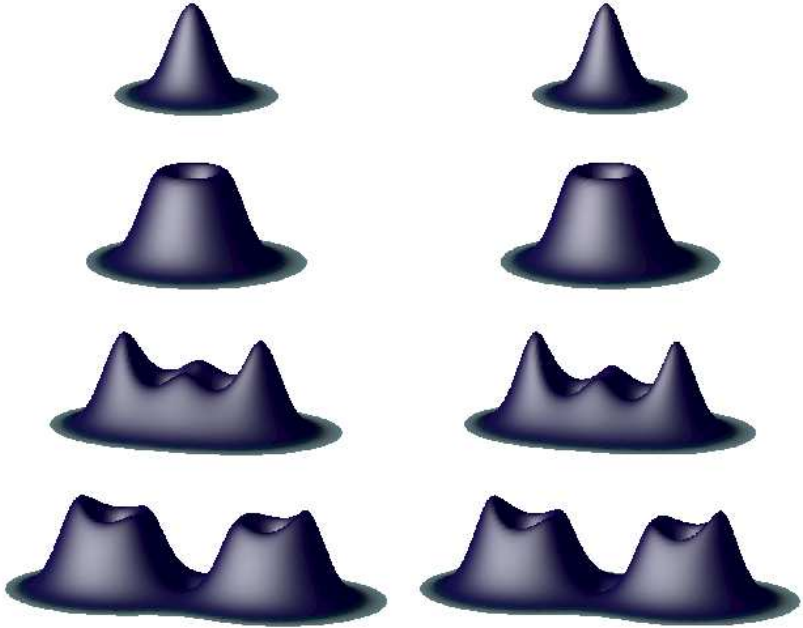}
\caption{
Plots of the topological charge density for soliton solutions
with $1\le B\le 4.$ The left-hand-side images are for the
baby Skyrme model and the right-hand-side images are for
the vector meson model. It is clear that there is a remarkable
similarity between the solitons in the two models. 
}
\label{fig-all4}
\end{center}
\end{figure}


In the last two rows of Table~\ref{tab-energy} 
we list the energies for the $B=3$ and
$B=4$ solitons in the two models, computed using the two-dimensional
code. The energies in this table confirm that all the solitons are
bound states. Even for the non-axial solitons, the energies in the two
models are very similar. In Figure~\ref{fig-all4} we present plots
of the topological charge density $B^0$ for $1\le B\le 4$ in both
models. The images on the left-hand-side correspond to the
baby Skyrme model and those on the right-hand-side to the vector meson model.
These plots confirm the amazing similarity between the solutions of the
two theories.

A modification of the baby Skyrme model was introduced in \cite{We},
in which the second term in (\ref{bslag}) is replaced by 
the symmetric mass term
\be
{\cal L}_{\rm sym}=-\frac{1}{2}m^2(1-\phi_3^2).
\label{smass}
\ee
As this term is symmetric under the replacement $\phi_3\mapsto-\phi_3,$
then the associated contribution to the energy density vanishes at
the centre of the soliton, where $\phi_3=-1.$ This contrasts with the
original asymmetric mass term in (\ref{bslag}), which gives an associated
maximal contribution to the energy density when $\phi_3=-1.$
Axially symmetric baby Skyrmions with $B>1$ have an energy density that
is maximal on a circle, and may be viewed as circular domain walls
separating the vacuum $\phi_3=1$ outside the wall from the
vacuum $\phi_3=-1$ inside the wall. For $B>2$ such configurations
are not favourable with the asymmetric mass term, because this results
in a contribution to the energy that is proportional to the area enclosed
within the wall. However, for the symmetric mass term (\ref{smass})
this contribution is removed and it turns out that the axially symmetric 
soliton has minimal energy for all values of $B$ \cite{We}. 
A similar phenomenon occurs in the (3+1)-dimensional Skyrme model,
where Skyrmions form hollow polyhedral shells without a mass term \cite{BS3},
but an asymmetric mass term produces more compact configurations 
that are no longer shell-like \cite{BS1,BMS}.    

To further examine the similarity between the Skyrme term and vector
meson stabilization, we compute axially symmetric solitons in the baby Skyrme
and vector meson models with the asymmetric mass term replaced
by the symmetric mass term (\ref{smass}) in both theories. 
The corresponding energies for solitons with $1\le B\le 4$ are
presented for both theories in Table~\ref{tab-axial}. Note that even
though we retain the value $g=20.83$, rather than fixing $g$ by equating 
the single soliton energies in the two theories, all the soliton
energies are again very close. In particular, the vector meson model
has axially symmetric bound states, sharing the same properties as
the baby Skyrme model with the symmetric mass term.   

\begin{table}[ht]
\centering
\begin{tabular}{|c|c|c|}
\hline
$B$ & $E^{{\rm BS}}$ & $E^{{\rm VM}}$ \\
\hline
1 & 18.18 & 18.30\\
2 & 32.91 & 33.28\\
3 & 48.29 & 48.82\\
4 & 63.89 & 64.58\\
\hline
\end{tabular}
\caption{
Energies of the (axially symmetric) 
minimal energy solitons with $1\le B\le 4,\ $ using
the symmetric mass term. 
$E^{{\rm BS}}$ is the energy in the baby Skyrme model 
and $E^{{\rm VM}}$ is the energy in the vector meson model.
}
\label{tab-axial}
\end{table}

\section{Conclusion}\news
We have presented the results of numerical computations of multi-solitons
in an alternative to the baby Skyrme model, in which the Skyrme term
is removed and stabilization is provided by the coupling to a massive
vector meson field. These results reveal a remarkable similarity between
the solitons of the two theories, which can be understood using a simple
derivative expansion. This is evidence for a similar correspondence
in the (3+1)-dimensional Skyrme model, which has been studied recently
using approximate methods, but has so far not received the detailed
numerical investigation that we have been able to perform here in this
lower-dimensional analogue.

There are several possible avenues for further investigation of the vector
meson theory introduced in this paper. The vector meson field introduced
here is the lower-dimensional analogue of the $\omega$ meson, so a 
natural extension is to include the lower-dimensional analogue of the 
$\rho$ meson. Also, all studies in the present paper have been
concerned with static solitons, and it would be interesting to study
soliton dynamics in this theory and make a comparison with dynamics
in the baby Skyrme model.

\section*{Acknowledgements}
DF thanks the STFC for a research studentship.
PMS thanks the STFC for support under the rolling grant ST/G000433/1.
The numerical computations were performed on the Durham HPC cluster Hamilton.

\end{document}